# Room-temperature operation of a molecular spin photovoltaic device on a transparent substrate

*Kaushik Bairagi\*, David Garcia Romero, Francesco Calavalle, Sara Catalano, Elisabetta Zuccatti, Roger Llopis, Fèlix Casanova, Luis E. Hueso\**

Dr. Kaushik Bairagi, David Garcia Romero, Francesco Calavalle, Dr. Sara Catalano, Elisabetta Zuccatti, Roger Llopis, Prof. Fèlix Casanova, Prof. Luis E. Hueso

CIC nanoGUNE, 20018 Donostia-San Sebastian, Basque Country, Spain
E-mail: k.bairagi@nanogune.eu; l.hueso@nanogune.eu

Prof. Fèlix Casanova, Prof. Luis E. Hueso
IKERBASQUE, Basque Foundation for Science, 48013 Bilbao, Basque Country, Spain

**Abstract:**

Incorporating multifunctionality along with the spin-related phenomenon in a single device is of great interest for the development of next generation spintronic devices. One of these challenges is to couple the photo-response of the device together with its magneto-response to exploit the multifunctional operation at room temperature. Here, the multifunctional operation of a single layer p-type molecular spin valve is presented, where the device shows a photovoltaic effect at the room temperature on a transparent glass substrate. The generated photovoltage is almost three times larger than the applied bias to the device which facilitates the modulation of the magnetic response of the device both with bias and light. It is observed that the photovoltage modulation with light and magnetic field is linear with the light intensity. The device shows an increase in power conversion efficiency under magnetic field, an ability to invert the current with magnetic field and under certain conditions it can act as a spin-photodetector with zero power consumption in the standby mode. The room temperature exploitation of the interplay among light, bias and magnetic field in the single device with a p-type molecule opens a way towards more complex and efficient operation of a complete spin-photovoltaic cell.

Vertical spin valve (SV) structures are widely used to study the spin transport through organic semiconductors (OSCs) [1–3]. Unlike inorganic semiconductors, OSCs are composed of light chemical elements resulting in a weak spin-orbit coupling and hyperfine interaction, which leads to long spin diffusion times[4–8]. Recently, and in spite of the relatively low carrier mobility of the materials studied, long distance spin transport has been achieved in several organic semiconductor based spin valves[3,9]. OSCs also offer a large degree of tunability in their mechanical, electrical and optical properties over their inorganic counterparts, which can be used to add multifunctionality in a single device[10–17]. One such example is the observation of a photovoltaic effect in an electron transporting (n-type) $C_{60}$ based spin valve, where the magnetic and optical response are coupled and that leads to several novel functionalities in a single device at low temperature[16]. However, the basic unit of a high performance small molecule organic photovoltaic cell is composed of an electron transporting (n-type) and a hole transporting (p-type) layer[18–20]. Thus, in order to achieve high photovoltaic response of the SV device towards its room temperature multifunctional operation, devices with only a hole transporting material has to be realized as a necessary step.

Here we present the room temperature operation of a p-type molecular spin photovoltaic (MSP) device on a transparent glass substrate. Our device has a simple vertical SV geometry composed of a bottom ferromagnetic (FM) electrode cobalt (Co) capped by a leaky $AlO_x$ barrier, a p-type organic molecule hydrogen phthalocyanine ($H_2Pc$) acting as the spin transporting channel and a top FM electrode $Ni_{80}Fe_{20}$ [2,3]. $H_2Pc$ is a commonly used p-type material for small molecule organic photovoltaic cells and is also known for its stability in ambient conditions[20–22]. **Figure 1a** presents the schematic diagram of an $H_2Pc$ molecule and Figure 1b presents the transfer characteristic of a 30 nm thick $H_2Pc$ based lateral organic field effect transistor (LOFET) respectively, which indicates the p-type operation with a lateral hole mobility of $3.3\times10^{-5}$ $cm^2V^{-1}s^{-1}$ [23]. The growth of the $H_2Pc$ molecule on the glass substrate is highly amorphous thus it exhibits isotropic charge transport and we do not expect large changes in the charge carrier mobility in a vertical geometry (See Figure S1). Figure 1c presents the schematic of the complete vertical SV device and Figure 1d represents the rigid band energy diagram of the device. The Fermi energy ($E_F$) of both the FM electrodes is well matched with the highest occupied molecular orbital (HOMO) of $H_2Pc$ with a hole injection barrier of 0.22 eV while the energy barrier for the electron injection into the lowest unoccupied molecular orbital (LUMO) is around 1.5 eV [24]. Hence, in our SV device, the charge carrier transport is most likely to be dominated by holes.

**Figure 2a** represents the current-voltage (I-V) response of the device in both dark and light irradiation conditions where the voltage bias is applied to the top $Ni_{80}Fe_{20}$ electrode and the bottom Co electrode is grounded. The device shows a photovoltaic response with a short circuit current ($I_{SC}$) of 3.2 nA and an open circuit voltage ($V_{OC}$) of 27.2 mV at room temperature for an incident light irradiation of 7.5 $mW.cm^{-2}$ when no external magnetic field is applied to the device. The values of $I_{SC}$ and $V_{OC}$ are relatively small as compared to other conventional photovoltaic cells since in our case the device is composed of a single photoactive layer[14,15,25–28]. Under an applied external magnetic field, the photovoltaic response of the device increases yielding a higher $V_{OC}$ value when the orientation of the FM electrodes are antiparallel as shown in Figure 2b. The power conversion efficiency increases from $1.7\times10^{-4}$% to $1.9\times10^{-4}$% (an increase of ≈12%) when the relative orientation of the FM electrodes changes from parallel to antiparallel state [16]. The mechanism of the photovoltaic effect under the absence of applied magnetic field and for parallel and anti-parallel orientation of the electrodes under magnetic field is schematized in Figure 2c-2d in the rigid band approximation at equilibrium. We infer from the I-V response of our device that, at equilibrium, there is an interface dipole formed at the $H_2Pc$/ $Ni_{80}Fe_{20}$ interface despite of the apparently good energy level alignment, however the bottom interface is less affected because of the presence of the $AlO_x$ layer which isolates the molecule from highly reactive Co electrode underneath. In the first case (Figure 2c), the photo-generated holes at the molecule-$Ni_{80}Fe_{20}$ interface are collected by the $Ni_{80}Fe_{20}$ electrode due to the in-built potential, before recombining with the electrons. The electrons then can either be transported to the bottom Co electrode by the in-built potential or lost due to partial recombination. Since $H_2Pc$ is a p-type organic semiconductor, the generated photocurrent is mainly dominated by the collected holes at the molecule-$Ni_{80}Fe_{20}$ interface. In the open-circuit mode, the injected spin-polarized carriers must compensate the photogenerated carriers or in other words, the applied bias must be equal and opposite to the photo-generated voltage. In the parallel configuration of the electrodes, the photo-generated voltage ($V_{OC, P}$) is the same as in the case without applied magnetic field while in the anti-parallel case, the injected spin-polarized holes are reflected by the bottom Co electrodes and can only be compensated by an enhanced photo-voltage $V_{OC,AP}$ (Figure 2d and 2e).

We define this change in open-circuit voltage ($\Delta V_{OC}$) as $V_{OC, AP} - V_{OC, P}$. **Figure 3a** presents the modulation of $\Delta V_{OC}$ with respect to magnetic field and light. At higher light intensities, the photo-generated carriers can be compensated with the application of a higher device bias in the parallel configuration and hence even a higher applied bias is needed for the compensation in the antiparallel configuration. We observe that the value of $\Delta V_{OC}$ increases linearly with increasing light intensity (Figure 3b). This can be useful for a spin-photodetector with zero power consumption since in the open-circuit mode of operation the current through the device is zero [29–31]. In the short circuit-mode, the photogenerated current remains constant over the range of applied magnetic field simply because the photogenerated carriers are not spin-polarized (Figure 2c).

In the dark conditions, we define magnetocurrent (MC) (in percentage) as $(I_P - I_{AP})/I_{AP} \times 100\ \%$, where $I_P$ and $I_{AP}$ are the currents in the parallel and anti-parallel configuration of the electrodes respectively. We observe an MC of 7% for an applied bias of 10 mV at room temperature (Figure 2d). Since the photo-generated voltage (27.2 mV) with a light irradiation of 7.5 mW.cm$^{-2}$ is larger than the applied bias (10 mV) to the device thus in these conditions the output current of the device can be modulated on both side of the zero current level by varying the light intensity. The light modulation of the output current does not affect the spin-polarized charge transport through the device as the photogenerated carriers are non-spin polarized. The photo-current only shifts the baseline of the magneto current response of the device. **Figure 4a** shows the output current vs magnetic field (I-B) response of the device under various light irradiation conditions at an applied bias of 10 mV. The overall device current decreases with increasing light intensity. The I-B responses of the device passes from an overall positive current to an overall negative current through a point where for a particular light intensity the anti-parallel state current can be set to zero. Similarly, for a slightly higher intensity, the parallel state current through the device can be set to zero. This effect could be used to realize a switch having zero power consumption in the standby state. Moreover, at yet another light intensity, the parallel state current can exactly be set equal and opposite of the anti-parallel state current ($I_P = - I_{AP}$) (See Figure S2). This can act as a magnetic current converter. However, in all light intensities the value of $\Delta I = I_P - I_{AP}$ remains the same for the applied bias of 10 mV.

We now focus on a fixed light intensity of 7.5 mW.cm$^{-2}$. In this case, the I-B response of the device can be modulated on both side of zero current level with a modified $\Delta I$ under different applied bias (Figure 4b). The modification of ΔI with both light and bias is summarized in Figure 4c. ΔI remains constant with increasing light intensity for a fixed device bias. This again confirms that the photo-generated carriers do not affect the spin-transport properties through the device. At a fixed bias the amount of spin-polarized charge carriers flowing through the device remains constant whereas for a change in applied bias, this amount changes owing to the change in ΔI. The current level in the parallel state of the device can be tuned both by applied bias and light. The applied bias to the device injects spin-polarized carriers into the molecular layer through molecule/ $Ni_{80}Fe_{20}$ interface whereas the incident light creates non-spin polarized carriers responsible for photocurrent generation. The interplay between light and applied biases can lead to certain states where for an applied bias equal to the open-circuit voltage, the parallel state current can be set to zero. At those particular light intensities and applied biases, spin-polarized current can be generated for the anti-parallel orientation of the electrodes as shown in Figure 4d. This complex electro-optical modulation of the device can lead to the operation of a spin-valve that consumes negligible power in the stand-by state and can act as a spin-photodetector [16,32,33].

In conclusion, we have illustrated the room temperature operation of a spin-photovoltaic device based on a thin film of hole transporting small molecule $H_2Pc$ on a transparent glass substrate. The photovoltaic power conversion efficiency was enhanced as a result of the enhanced photovoltage generation with the application of a small applied magnetic field. The device can act as a perfect magnetic current converter, as a spin-photodetector and can generate purely spin-polarized current under certain applied bias and light irradiation conditions. Our approach opens the door towards the generation of a more complex and highly efficient spin-photovoltaic device, e.g. a p-n heterojunction device, which is a necessary building block for future devices operating at room temperature on transparent substrates.

**Experimental Section**

*Device fabrication:*

Vertical spin valves with Co/AlO$_x$/H$_2$Pc/Ni$_{80}$Fe$_{20}$ have been fabricated *in-situ* in an ultrahigh vacuum (UHV) chamber with base pressure less than 10$^{-9}$ mbar. The pyrex glass substrates were cleaned in a ultrasonic bath using acetone and isopropanol subsequently and then dried with a N$_2$ flow. The substrates were cooled down to liquid N$_2$ temperature before the deposition of the materials. At first, eight 12 nm thick Co lines were deposited through shadow masking technique to define the bottom electrodes. A 1.5 nm Al layer was then deposited everywhere on the sample and semi-oxidized to form a leaky AlO$_x$ barrier. A 90 nm thick H$_2$Pc was then deposited again with shadow masking technique to form the molecular layer on Co/AlO$_x$ while leaving one junction without the molecular layer as reference. Finally, the top Ni$_{80}$Fe$_{20}$ lines were deposited using another shadow mask to complete the device. The area of the cross-bar geometry devices were ranging from 210×280 µm$^2$ to 370×550 µm$^2$. Co, Ni$_{80}$Fe$_{20}$ and Al were purchased from Lesker (purity: 99.95%) and were used as received. Co, Ni$_{80}$Fe$_{20}$ were evaporated from an e-beam evaporator in one of the UHV chamber with a rate of 1 Å/s (for the top Ni$_{80}$Fe$_{20}$ deposition, the starting rate was 0.1 Å/s for the first 2 nm to protect the soft organic layer). Al was thermally evaporated from a Knudsen cell at a rate of 1 Å/s. H$_2$Pc was purchased from Sigma Aldrich (with 99.99% purity) and was used without further purification. The molecules were evaporated at a rate of 0.1 Å/s from another Knudsen cell from a separate chamber.

*Thin film characterization and electrical measurements:*

Both the metal and the organic materials were calibrated using the quartz crystal monitor during the evaporation. The film thicknesses were measured using X-ray reflectivity (XRR) technique and the morphologies were checked by atomic force microscopy (AFM) technique.

The devices were measured in a standard four-probe configuration in a variable temperature Lakeshore probe station (equipped with magnetic field) under high vacuum.

**Acknowledgements**

This work was supported by the Spanish MICINN under the Maria de Maeztu Units of Excellence Programme (MDM-2016-0618) and under Projects MAT2015-65159-R and RTI2018-094861-B-100, and by the European Union H2020 under the Marie Curie Actions (796817-ARTEMIS and 766025-QuESTech).

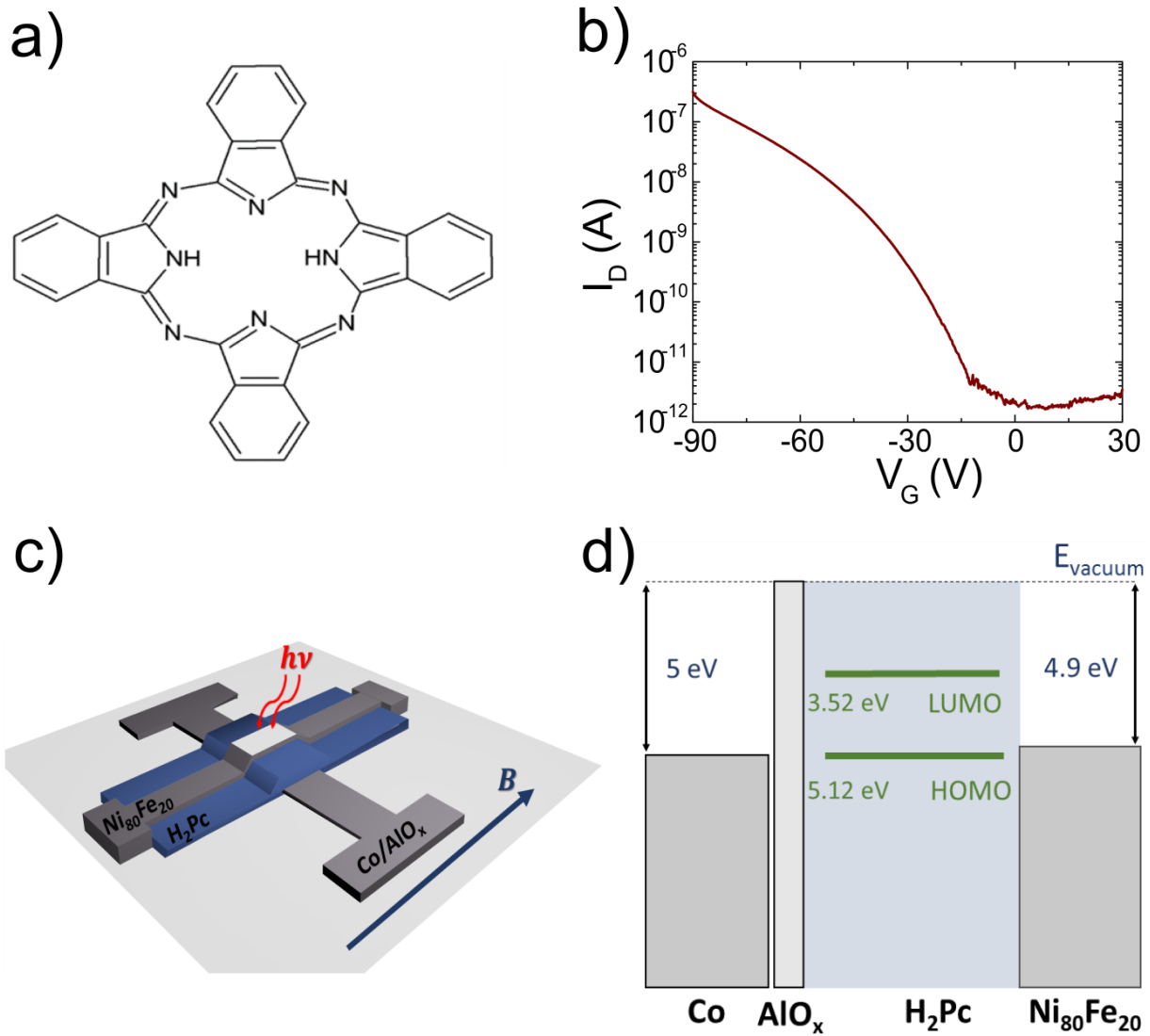

**Figure 1.** a) Schematic of hydrogen phthalocyanine (H$_2$Pc) molecule. b) Transfer characteristic of a 30nm thick H$_2$Pc based lateral organic field effect transistor (L = 10 μm; W = 1 mm) showing p-type operation with a source-drain bias V$_D$ = 40V. c) Schematic presentation of the H$_2$Pc based spin-valve operating as a spin-photovoltaic device on a glass substrate. d) Rigid band energy diagram of the device. The Fermi level (E$_F$) of both Co and Ni$_{80}$Fe$_{20}$ electrodes matches very well to the HOMO level of the H$_2$Pc molecule, favoring a hole dominated transport through the device.

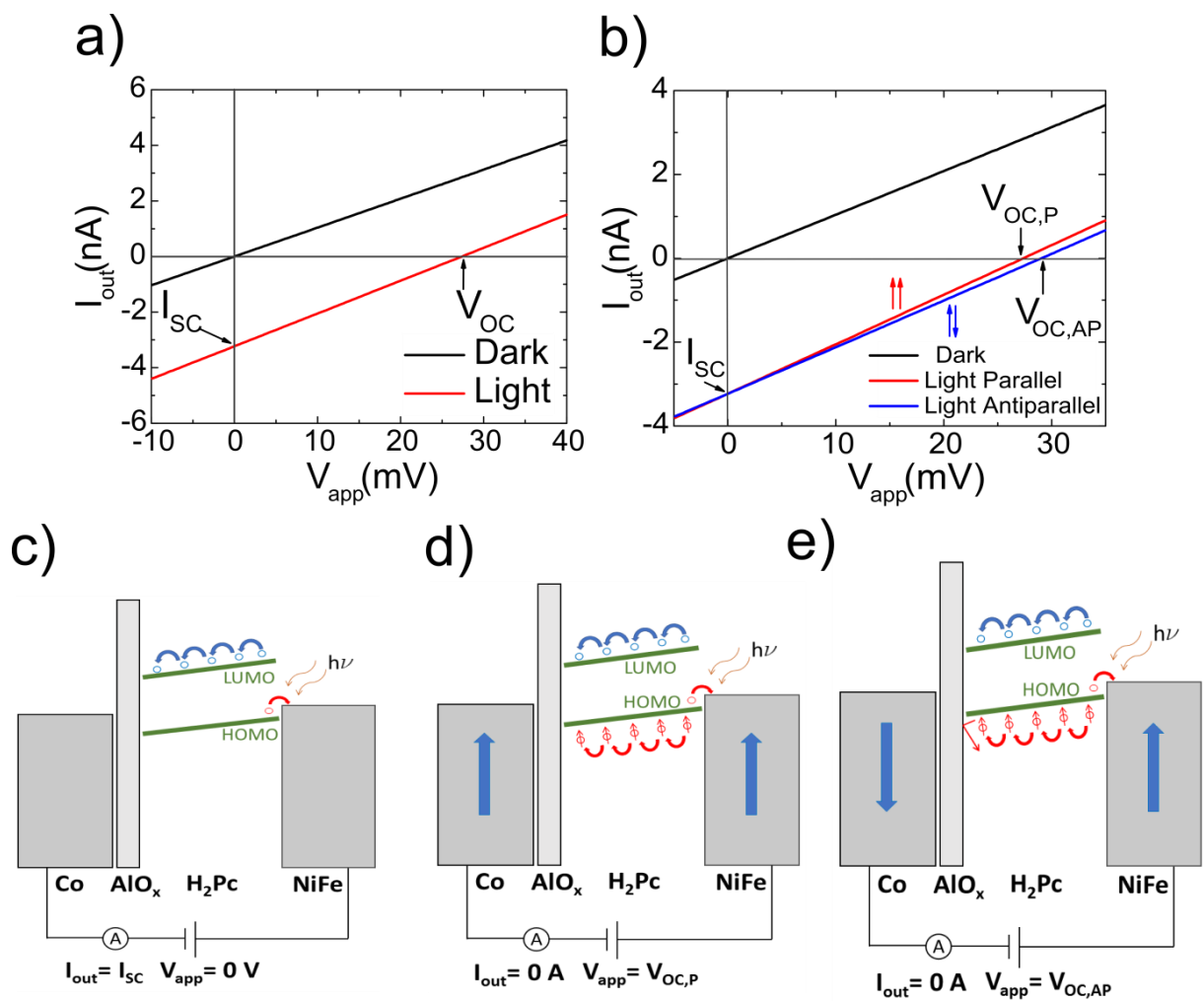

**Figure 2.** a) Current-voltage (I-V) response of the MSP device in dark and light (7.5 mW.cm$^{-2}$) irradiation conditions without the application of any magnetic field. b) I-V response of the device in dark and during light irradiation when the electrodes are in parallel or anti-parallel configuration. c) Mechanism of electron-hole transport upon light irradiation under zero bias conditions. d)-e) Spin dependent charge carrier transport through the molecular layer under light irradiation in both parallel and anti-parallel configurations of the electrodes in the open circuit mode.

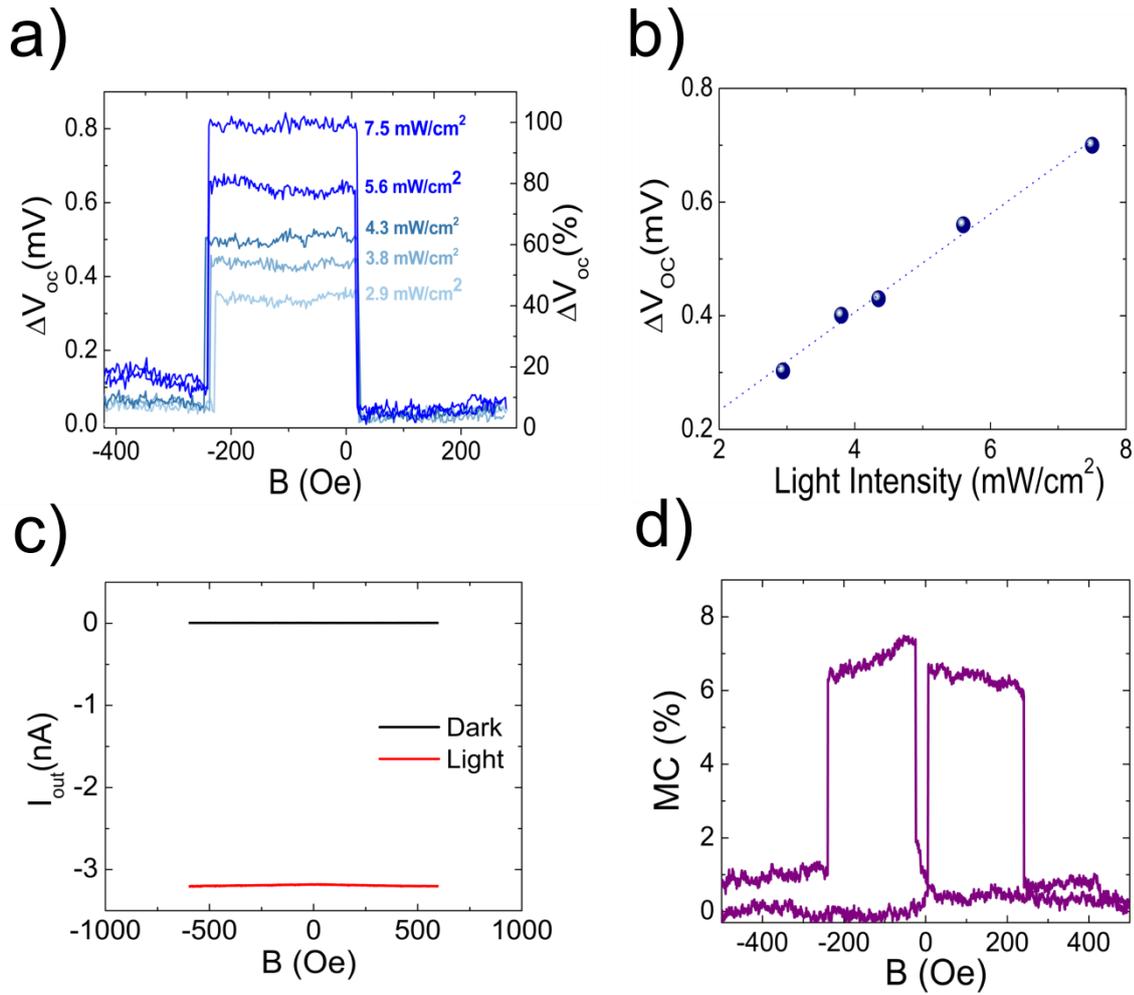

**Figure 3.** a) Modulation of the open-circuit voltage with magnetic field at various light intensities. The net current through the device is zero. b) The difference in open-circuit voltage ($\Delta V_{OC} = V_{OC, AP} - V_{OC, P}$) as a function of light intensities. $\Delta V_{OC}$ is linear with respect to the light intensities. c) I-B response of the device with zero applied bias in dark and light irradiation conditions. The photo-generated current is unaffected on the application of magnetic field as it is not spin polarized. d) Magnetocurrent (MC) response of the device at 300K.

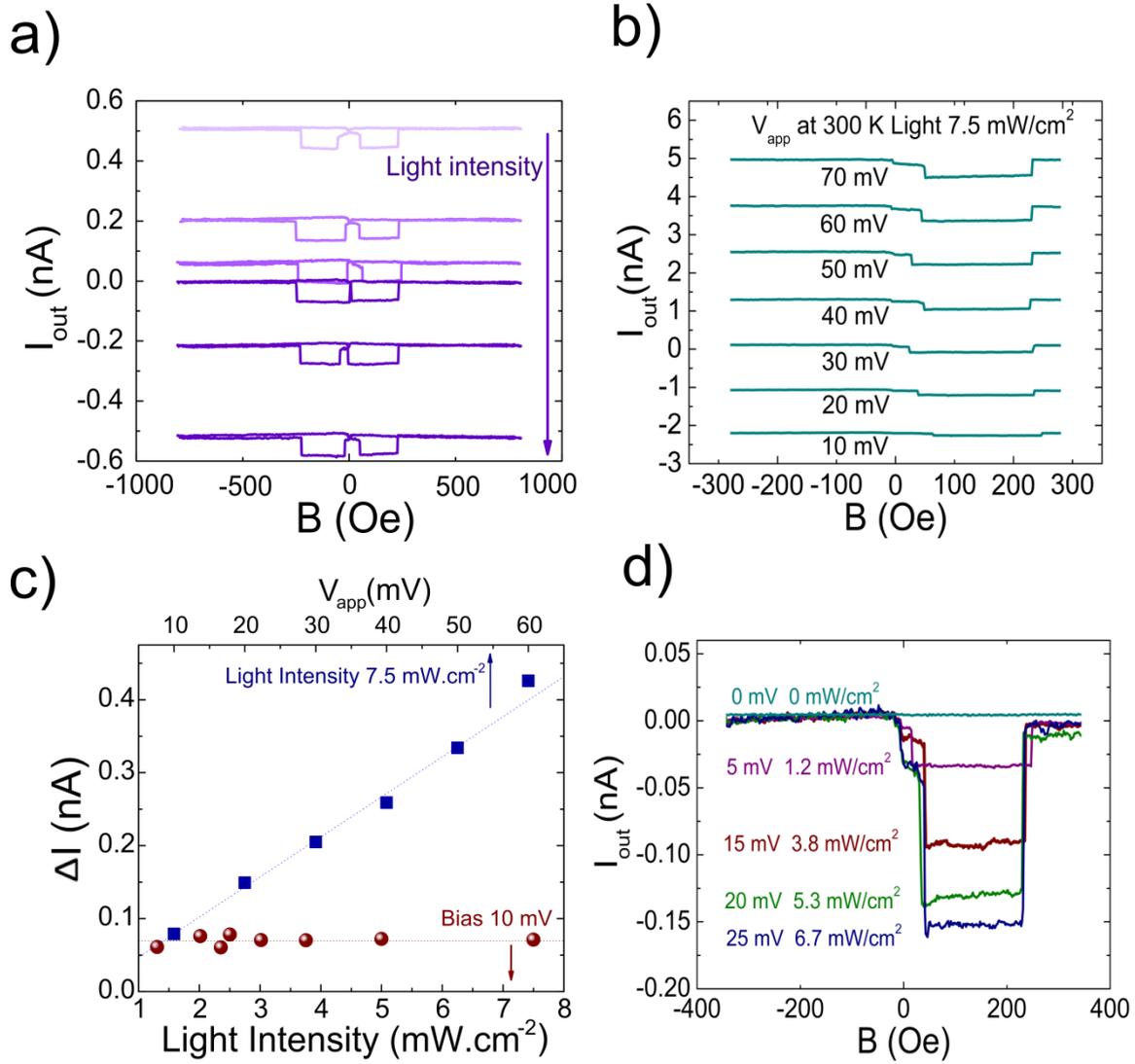

**Figure 4.** a) Modulation of output current of the device as a function of the magnetic field with varying light intensities at a constant bias of 10 mV. b) I-B responses of the device under constant light irradiation but varying the bias from 10 mV to 70 mV. c) The difference in current in parallel and anti-parallel state (ΔI) as a function of light intensity for a constant bias of 10 mV and as a function of applied bias for a constant light intensity of 7.5 mW.cm$^{-2}$. d) Electro-optical modulation of the device under varying applied bias and light intensity to have zero output current in the parallel state, meaning that the generated open-circuit voltage cancels exactly the applied bias.

Supporting Information

**Room-temperature operation of a molecular spin photovoltaic device on a transparent substrate**

*Kaushik Bairagi\*, David Garcia Romero, Francesco Calavalle, Sara Catalano, Elisabetta Zuccatti, Roger Llopis, Fèlix Casanova, Luis E. Hueso\**

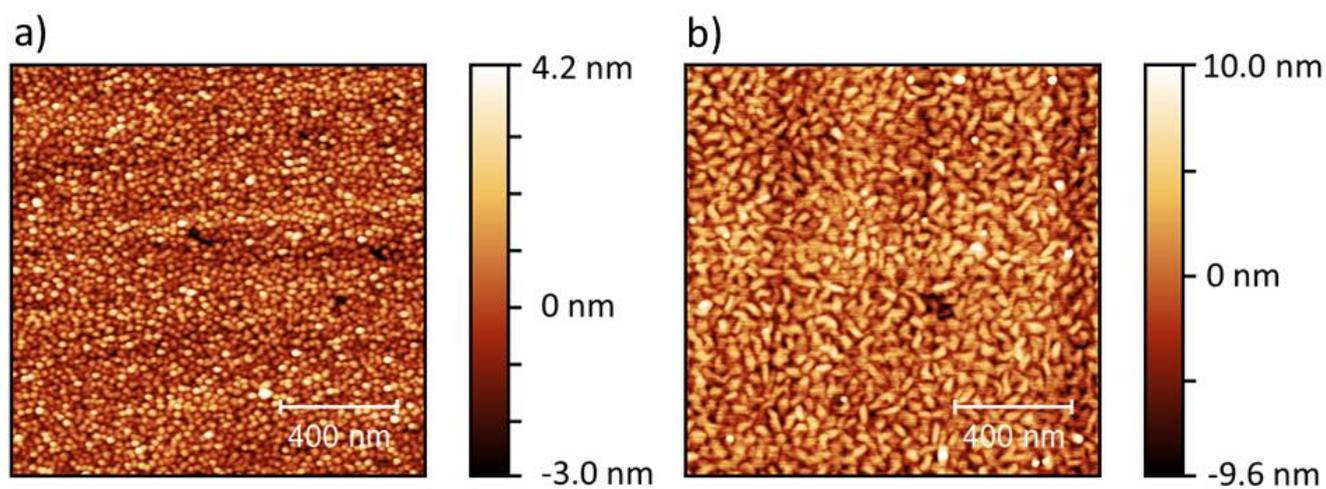

**Figure S1.** AFM image of 30 nm $H_2Pc$ grown on glass substrate with a) LT condition having rms roughness of 1.9 nm b) RT condition having rms roughness of 2.4 nm

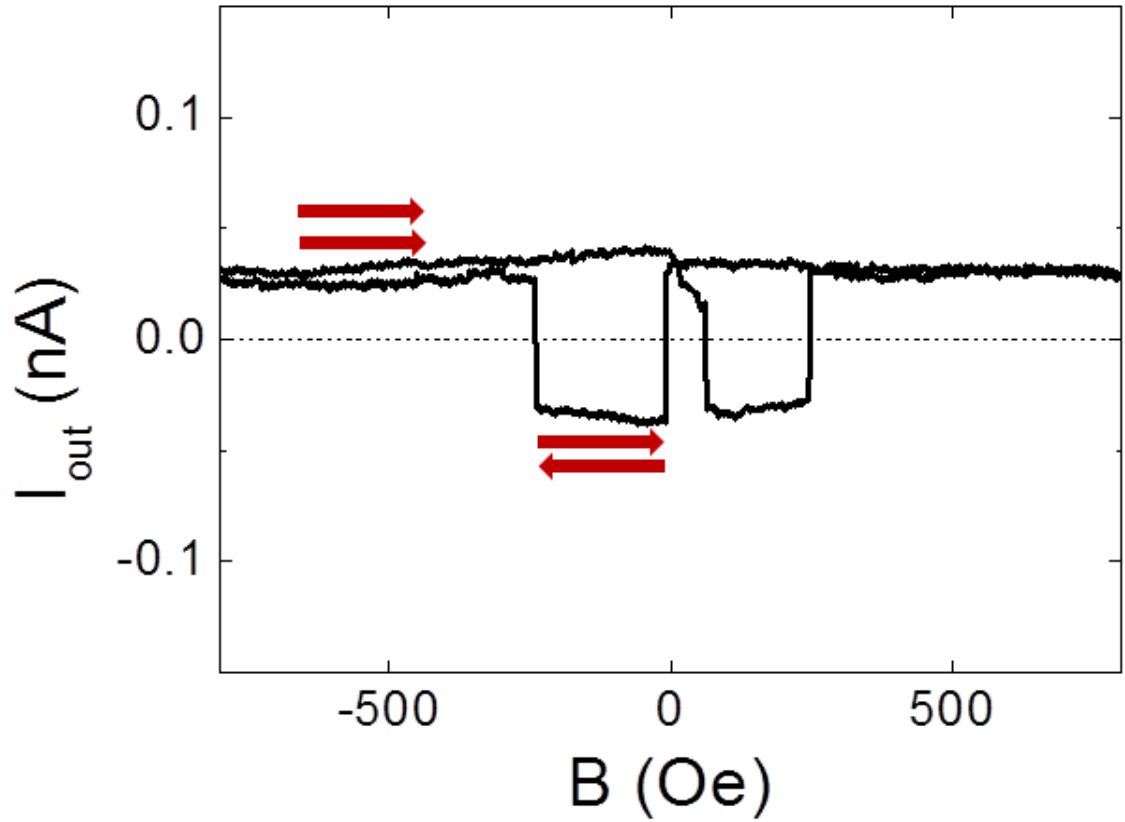

**Figure S2.** Light modulation of the magnetocurrent leads to a state where the parallel state current is equal and opposite to the anti-parallel state current (device bias: 10 mV; light intensity: 2.5 mW.cm$^{-2}$).

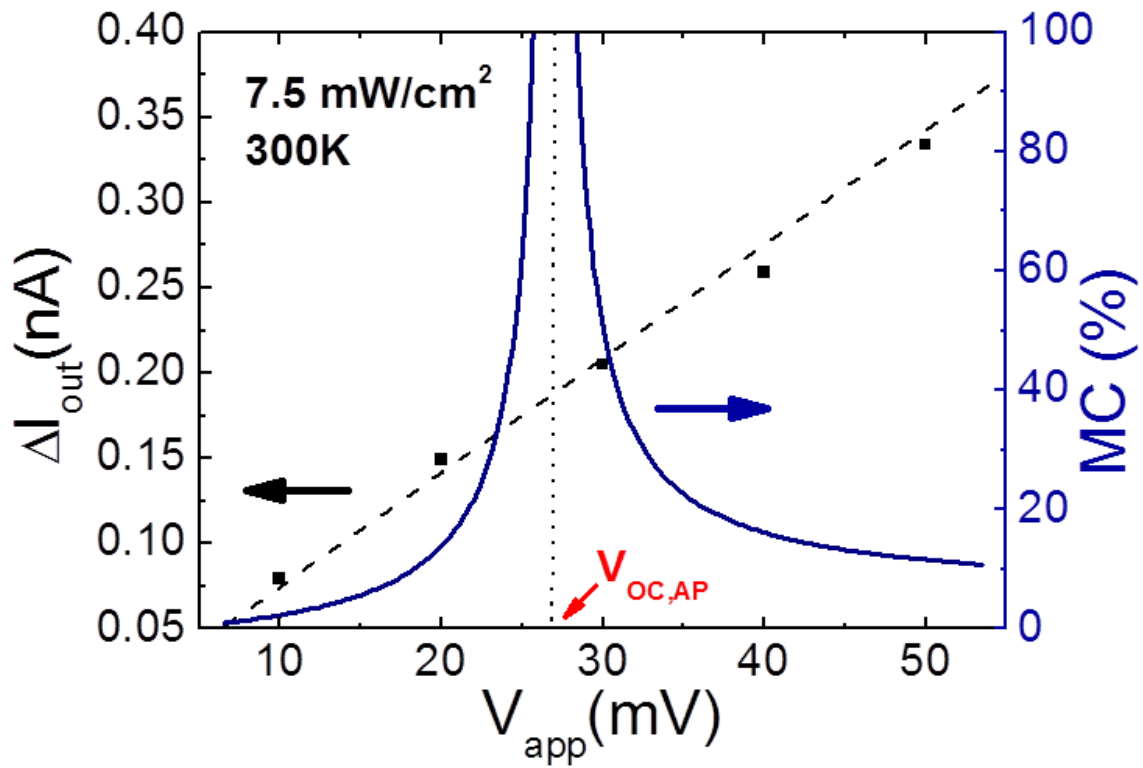

**Figure S3.** Difference in current in parallel and anti-parallel state (ΔI) and the corresponding MC as a function of the applied bias ($V_{app}$) (calculated from the linear dependence of ΔI as a function of $V_{app}$) under light irradiation of 7.5 mW.cm$^{-2}$. At an applied bias equal to the open circuit voltage, the MC value tends to infinite making it favorable for a light sensor.